\documentclass[english,a4paper,pre,twocolumn,amsmath,amssymb,longbibliography]{revtex4-1}
\usepackage{graphicx,color}
\usepackage{bm}
\usepackage{color}
\usepackage{soul}
\usepackage[colorlinks,linkcolor=magenta,citecolor=magenta]{hyperref}

\newcommand{\dd}{\mathrm{d}}

\newcommand{\mean}[1]{\langle #1 \rangle}
\newcommand{\Int}[1]{\int\dd #1\;}
\newcommand{\IInt}[3]{\int_{#2}^{#3}\dd #1\;}

\renewcommand{\vec}[1]{\mathbf #1}

\newcommand{\al}{\alpha}
\newcommand{\gam}{\gamma}
\newcommand{\eps}{\varepsilon}
\newcommand{\kap}{\kappa}

\newcommand{\sig}{\sigma}
\newcommand{\om}{\omega}
\newcommand{\Om}{\Omega}

\newcommand{\im}{\mathrm i}
\newcommand{\kT}{k_\text{B}T}
\newcommand{\kTi}{k_\text{B}\tilde T}
\newcommand{\To}{T_\text{o}}
\newcommand{\Tg}{T_\text{g}}
\newcommand{\df}{d_\text{f}}

\begin{document}

\title{Modeling non-linear dielectric susceptibilities of supercooled molecular liquids}

\author{Thomas Speck}
\affiliation{Institut f\"ur Physik, Johannes Gutenberg-Universit\"at Mainz, Staudingerweg 7-9, 55128 Mainz, Germany}

\begin{abstract}
  Advances in high-precision dielectric spectroscopy has enabled access to non-linear susceptibilities of polar molecular liquids. The observed non-monotonic behavior has been claimed to provide strong support for theories of dynamic arrest based on thermodynamic amorphous order. Here we approach this question from the perspective of dynamic facilitation, an alternative view focusing on emergent kinetic constraints underlying the dynamic arrest of a liquid approaching its glass transition. We derive explicit expressions for the frequency-dependent higher-order dielectric susceptibilities exhibiting a non-monotonic shape, the height of which increases as temperature is lowered. We demonstrate excellent agreement with the experimental data for glycerol, challenging the idea that non-linear response functions reveal correlated relaxation in supercooled liquids.
\end{abstract}

\maketitle


\section{Introduction}

\begin{figure}[b!]
  \centering
  \includegraphics{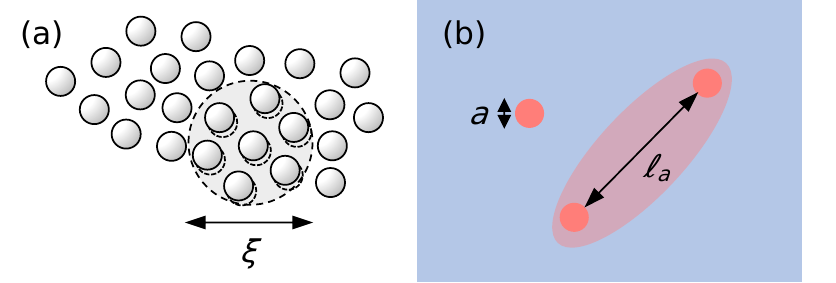}
  \caption{Competing qualitative views on the relaxation mechanism in supercooled liquids: (a)~Cooperative motion of particles captured by a static length scale $\xi$. (b)~Localized active regions (red) with size $a$ in a sea of jammed immobile particles (blue). To relax, two excitations with typical distance $\ell_a$ need to connect through a larger excited active region (shaded).}
  \label{fig:comp}
\end{figure}

Dynamic arrest is a generic phenomenon found in a wide range of materials such as liquids (including water~\cite{angell08,limm14}), gels~\cite{zacc07,lu08}, and even organic tissue~\cite{bi16}. Discerning competing theoretical perspectives on the mechanism underlying the kinetic arrest of supercooled liquids approaching the glass transition is a persistent challenge~\cite{biroli13,royall18,royall20}. The major obstacle is arguably the absence of a distinct structural change accompanying the dramatic increase of structural relaxation time over more than ten orders of magnitude within a rather small temperature range. One route to resolve this conundrum is through assuming \emph{cooperativity}, \emph{i.e.}, more and more particles (atoms, molecules, or colloidal particles) have to move in a correlated fashion to relax a portion of the liquid [Fig.~\ref{fig:comp}(a)]. This is the mechanism put forward by random first-order theory (RFOT)~\cite{lubc07,bouchaud04} based on earlier arguments by Adam and Gibbs~\cite{adam65}. It implies a static correlation length $\xi$ quantifying the range of ``amorphous order'' that increases as the glass transition is approached~\cite{karmakar14,hallett18}.

Alternatively, dynamic facilitation ascribes dynamic arrest to the emergence of kinetic constraints~\cite{chan10,garrahan18,speck19,katira19}. It posits that relaxation occurs through localized excitations, \emph{active regions} that can sustain particle motion, which facilitate the motion in neighboring regions in a hierarchical manner~\cite{palm84}. The structural relaxation time is determined by the density of these excitations, which plunges as the temperature is reduced below an onset temperature while the excitations themselves remain small (tens of particles in model glass formers) and spatially independent~\cite{keys11}. Their dynamics, however, is assumed to be strongly correlated, which is sufficient to obtain a super-Arrhenius increase of the relaxation time (Sec.~\ref{sec:fac}). The relevant length scale thus is not a correlation length $\xi$ but the typical distance $\ell_a$ between these excitations, which move further and further apart as the temperature is reduced [Fig.~\ref{fig:comp}(b)]. Originally, dynamic facilitation did not define these excitations but focuses on their statistical description. What has been criticized as weakness can also be seen as a strength since it allows to model the elementary building blocks of dynamic arrest independent of microscopic details (but see Refs.~\citenum{haysim21,ortlieb21} for recent progress towards a construction of excitations and also Ref.~\citenum{scho16} for ``learning'' the structural signature of excitations). Strong support for dynamic facilitation comes from the prediction of a coexistence between active and inactive dynamic phases~\cite{mero05,garr09}, for which there is now ample numerical~\cite{hedges09,speck12,campo20} and even experimental evidence~\cite{pinch17,abou18}. While dynamic facilitation does not require static correlations to explain the basic facts of dynamic arrest, including structural correlations enriches the dynamic facilitation scenario with the possibility that the dynamic coexistence terminates at a lower critical point~\cite{elma10,turci17,royall20}. Other theoretical approaches that we will not discuss include mode-coupling theory~\cite{gotz92,janssen18} and elastic models~\cite{hecksher15}.

Both RFOT and dynamic facilitation predict a steep super-Arrhenius increase of the structural relaxation time, and the available data is insufficient to discriminate both approaches. Broadband dielectric spectroscopy gives access to an exceptionally wide range of time scales and allows to study the molecular dynamics of polar liquids~\cite{kremer,lunkenheimer02}. Going beyond the linear dielectric response is technically demanding but promises insights into higher-order correlations~\cite{lunkenheimer17,richert17}. The observed non-monotonic shape of these non-linear susceptibilities with a peak close to the frequency set by the relaxation time has been interpreted as evidence for ordered domains and growing spatial correlations~\cite{bouchaud05,crauste10,brun12}, and claimed to rule out dynamic approaches such as dynamic facilitation~\cite{albe16}.

At variance with this assertion, in Ref.~\citenum{speck19} it has been argued that dynamic facilitation allows for a collective response to an external field; and that the observed relation between the third and fifth-order peak susceptibilities is compatible with dynamic facilitation. The argument was based on a single dominant length scale. In a recent response~\cite{biroli21}, Biroli, Bouchaud, and Ladieu correctly point out that the temperature dependence of the corresponding static linear susceptibility $\chi_1(0)$ is at variance with the experimentally observed $\chi_1(0)\sim1/(\kT)$. Here we revisit the modeling of the dielectric response of a supercooled polar liquid from the perspective of dynamic facilitation and argue that it can account for the non-trivial response of a polar liquid. We focus on a coarse-grained polarization field $m_\parallel$ that determines the relaxation dynamics of the individual dipoles. Using the statistical arguments of dynamic facilitation and building onto the detailed insights into model glass formers obtained by Keys \emph{et al.}~\cite{keys11}, we derive an explicit expression for this field from which we extract (effective) susceptibilities (Sec.~\ref{sec:chi}). We then turn to the experimental results provided by Refs.~\citenum{crauste10,albe16} and demonstrate that our results are in agreement with the data (Sec.~\ref{sec:exp}).


\section{Modeling dielectric relaxation}

\subsection{Polarization density}

We are interested in polar liquids with molecular elementary dipoles $\mu_0$ in $d=2,3$ dimensions~\cite{flenner15}. In a liquid at zero external electric field $E=0$, the orientations of these dipoles are uniformly distributed and their average is thus zero. Applying a time-dependent electric field $\vec E(t)$ with frequency $\om$ along a fixed direction, dipoles try to align with the field. The induced polarization density in a small volume $a^d$ is
\begin{equation}
  \vec p(t) = \mu_0\Int{\Omega} \vec e\psi(\vec e,t),
\end{equation}
where the integral is over the $d$-dimensional unit sphere and $\vec e$ is the instantaneous unit orientation with probability density $\psi(\vec e,t)$.

For the dynamics of individual dipoles, we assume a simple rotational relaxation on the time scale $\tau_0$ leading to
\begin{equation}
  \tau_0\partial_t\vec p = \rho\vec h - \vec p
  \label{eq:p}
\end{equation}
with $\rho$ the number density of dipoles. Further assuming that the effective field $\vec h(t)=\vec m e^{\im\om t}$ oscillates with frequency $\om$ and has amplitude $\vec m$, the complex solution of Eq.~\eqref{eq:p} reads
\begin{equation}
  \vec p(\om) = \frac{\rho\vec m}{1+\im\om\tau_0}.
\end{equation}
The amplitude $\vec m$ is a coarse-grained field that is uniform over the microscopic length $a$ and depends on the field strength $E$ and the frequency $\om$. Following the experiments~\cite{crauste10,albe16}, we consider the modulus of the averaged polarization density
\begin{equation}
  P(\om) = |\mean{p_\parallel(\om)}| = \frac{\rho m_\parallel(\om)}{(1+\om^2\tau_\al^2)^{\beta/2}}.
  \label{eq:p:mod}
\end{equation}
Local interactions typically renormalize the relaxation time $\tau_\al$ and multiple time scales lead to an effective exponent $\beta$ different from unity. Here, $m_\parallel(\om)$ is the average amplitude parallel to the electric field. This is the central quantity that we will model using a statistical mechanics approach. Through the expansion ($\eps_0$ is the vacuum permittivity)
\begin{equation}
  P(\om) = \eps_0\sum_{k=1,3,\dots}\chi_k(\om)E^k
\end{equation}
in the field strength $E$, we define the effective susceptibilities $\chi_k(\om)$. These $\chi_k(\om)$ depend on a single frequency and serve as approximants for the experimentally measured moduli of susceptibilities. Taking into account the full time dependence of the electric field would lead to complex susceptibilities $\hat\chi_k(\om_1,\dots,\om_k)$ different from $\chi_k(\om)$. Calculating the modulus $|\hat\chi_k|$ explicitly for a class of stochastic models~\cite{diez12,diez17,diez18} yields expressions with a more complicated dependence on frequency compared to results based on Eq.~\eqref{eq:p:mod}. Here we focus on modeling $m_\parallel$ and content ourselves with an effective dynamics described by $\tau_\al$ and $\beta$.

\subsection{Independent dipoles: Debye relaxation}

Before we embark on discussing supercooled liquids, let us consider the simplest case in which all dipoles are independent and the amplitude is determined by thermal equilibrium. A single dipole $\mu_0\vec e$ with unit orientation $\vec e$ has the energy $\mathcal E_1=-\mu_0\vec e\cdot\vec E=-\mu_0E\cos\theta$ with $\theta$ the polar angle enclosed by $\vec E$ and $\vec e$. The equilibrium partition function reads
\begin{equation}
  Z_1 = \Int{\Om} \exp\left\{\tfrac{\mu_0E}{\kT}\cos\theta\right\} = F_d(\mu_0E/\kT),
\end{equation}
where $\theta$ is the polar angle between the field direction and the orientation $\vec e$. Here, $F_d(x)$ depends on the spatial dimension $d$ and reads
\begin{equation}
  F_2(x) = 2\pi I_0(x), \qquad
  F_3(x) = 4\pi\frac{\sinh x}{x}
\end{equation}
with $I_k(x)$ the modified Bessel functions of the first kind. The average moment along the field thus is
\begin{equation}
  m_\parallel^{(0)} = \mu_0\mean{\cos\theta} = \kT\partial_E\ln Z_1 = \mu_0f_d(\mu_0E/\kT)
  \label{eq:m0}
\end{equation}
with
\begin{equation}
  f_2(x) = \frac{I_1(x)}{I_0(x)}, \qquad f_3(x) = \coth x-\frac{1}{x}.
\end{equation}
For the modulus of the polarization density [Eq.~\eqref{eq:p:mod}], we obtain ($\beta=1$)
\begin{equation}
  P(\om) = \frac{\rho\mu_0f_d(\mu_0E/\kT)}{(1+\om^2\tau^2)^{1/2}}.
  \label{eq:P0}
\end{equation}
Expanding $f_d(x)\approx x/d$ to linear order of the field strength $E$ we thus recover Debye's seminal result
\begin{equation}
  \chi_1^{(0)}(\om) = \frac{\rho\mu_0^2/d}{\eps_0\kT(1+\om^2\tau^2)^{1/2}}
\end{equation}
for the modulus of the linear susceptibility.


\section{Dynamic facilitation}

\subsection{Background}
\label{sec:fac}

We briefly recapitulate the basic ideas of dynamic facilitation following Ref.~\citenum{speck19}. We assume that the instantaneous local state (over a length $a$) of the liquid can be described as either active, \emph{i.e.} supporting motion of its constituent molecules, or inactive. Active regions need to be thermally excited with energy $J(a)$ while inactive regions have lower entropy $\Delta s(a)$. The fraction of active regions on scale $a$ in thermal equilibrium thus becomes
\begin{equation}
  c(a,T) = \frac{e^{-J/\kT}}{e^{-J/\kT}+e^{-\Delta s/k_\text{B}}} = \frac{1}{1+e^{J/\kTi}}
  \label{eq:c}
\end{equation}
with $1/\tilde T\equiv1/T-1/\To$ and onset temperature $\To\equiv J/\Delta s$. Cooling below the onset temperature, the concentration of active regions starts to decline sharply. Note that $\To$ is assumed to be a material property and independent of $a$.

At the heart of dynamic facilitation is the idea that active regions, small thermally excited pockets of mobility in a sea of immobile particles, remain active and need to connect to other active regions to relax. The typical distance between active regions of size $a$ is $\ell_a/a\simeq[c(a,T)]^{-1/\df}$ with possibly fractal dimension $\df\leqslant d$ of the active regions. We posit that the dominant mechanism for relaxation is to wait for an active region on the scale $\ell_a$ that connects active regions. The barrier to overcome thus is $\Delta F/\kT=[J(\ell_a)-J(a)]/\kTi$ with waiting time
\begin{equation}
  \tau_1(a,T) = \tau_\ast e^{\Delta F/\kT},
  \label{eq:tau:a}
\end{equation}
where $\tau_\ast$ is a typical fast time scale of the liquid.

The final ingredient is the dependence of the energy $J(a)$ on length, for which self-similarity
implies
\begin{equation}
  J(a) = J(\sig)[1+\gam\ln(a/\sig)]
  \label{eq:J}
\end{equation}
using as reference length the linear extent $\sig$ (\emph{viz.} particle diameter) of molecules. Here, $\gam$ is a dimensionless material-dependent coefficient. We thus obtain
\begin{equation}
  J(\ell_a)-J(a) = \gam J(\sig)\ln(\ell_a/a) \simeq -\frac{\gam J(\sig)}{\df}\ln c(a,T).
\end{equation}
Now inserting $c\simeq e^{-J/\kTi}$ for $T\ll\To$, we obtain the structural relaxation time~\cite{elma09,elma10a,keys11}
\begin{equation}
  \tau_\al(T) = \tau_1(\sig,T) = \tau_\ast\exp\left\{\mathcal J^2\left(\frac{\To}{T}-1\right)^2\right\}
  \label{eq:tau:al}
\end{equation}
with $\mathcal J\equiv\sqrt{\gam/\df}J(\sig)/\kT_\text{o}$.

\subsection{Dielectric relaxation}
\label{sec:chi}

\begin{figure}[b!]
  \centering
  \includegraphics{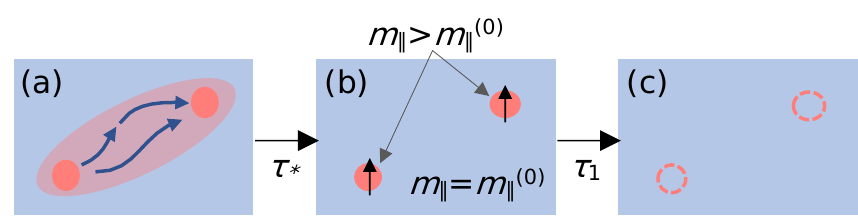}
  \caption{Caricature of the proposed mechanism: (a)~Surging, string-like displacements lead to highly correlated particle motion in an excited region over a short time $\tau_\ast$. (b)~The excited region leaves two active regions in which particle motion is possible, the local field $m_\parallel$ of which ``remembers'' the induced correlation. (c)~This correlation survives for a time determined by the relaxation time $\tau_1$ [Eq.~\eqref{eq:tau:a}].}
  \label{fig:sketch}
\end{figure}

Viewing small active regions as mobility defects, these defects are uncorrelated \emph{in space}. Due to the kinetic constraints, however, they are \emph{dynamically} correlated. Consider two active regions with extent $a$ connected through a short-lived (on the order of $\tau_\ast$) excited region with linear size $\ell_a$ [Fig.~\ref{fig:sketch}(a)]. Numerical~\cite{gebremichael04,bergroth05,keys11} and experimental~\cite{gokh14} evidence supports the picture of string-like surging motion, in which individual particles move only a short distance but these displacements are highly collective. Hence, particles that are initially uncorrelated become correlated over a length $\ell_a$ for a short time. The kinetic constraints imply that this induced correlation between the two active regions is preserved [Fig.~\ref{fig:sketch}(b)] until they are revisited by an excited region [Fig.~\ref{fig:sketch}(c)].

We now cast this qualitative argument into an expression for $m_\parallel$. Due to the collective motion, the electric field excites an additional contribution $m_\parallel(a)$ to the coarse-grained amplitude that governs the dipoles within the two active regions. To determine $m_\parallel(a)$, we consider the total dipole moment
\begin{equation}
  \bm\mu = \sum_{i=1}^n \mu_0\vec e_i
\end{equation}
of $n(a)\simeq(\ell_a/a)^{\df}$ independent dipoles in the excited region. We ask for the probability that the field excites the moment $\bm\mu$, the magnitude of which is bounded by
\begin{equation}
  |\bm\mu|^2 \leqslant \mu^2 = \overline{\bm\mu^2} = \mu_0^2\sum_{i,j=1}^n\overline{\vec e_i\cdot\vec e_j} = \mu_0^2n
\end{equation}
where the overline indicates the average over independent orientations with $\overline{\vec e_i\cdot\vec e_j}=\delta_{ij}$. The amplitude per dipole
\begin{equation}
  m_\parallel(a) = \frac{\kT}{n}\partial_E\ln Z_n = \frac{\mu}{n}f_d(\mu E/\kT)
  \label{eq:m:a}
\end{equation}
can now be found along the same lines as for the independent dipoles [Eq.~\eqref{eq:m0}] but with $\mu_0$ replaced by $\mu$. Note that we do not require the dipoles to be correlated and to respond rigidly, rather the local amplitude originating from a short correlated excitation remains elevated. In line with dynamic facilitation, this amplitude persists until the active regions are again visited by an excited region, which takes on average a time $\tau_1(a,T)$. Hence, the fraction of active regions of size $a$ possibly contributing to the dielectric response is
\begin{equation}
  \hat c(a,\om,T) = c(a,T)e^{-\kap[\om\tau_1(a,T)]^{-1}},
\end{equation}
where the second term is the probability that the dynamic correlation survives the time $t\simeq\kap/\om$ set by the oscillations of the external field.

The total amplitude is modeled as
\begin{equation}
  m_\parallel(\om) = m_\parallel^{(0)} + \frac{1}{N}\IInt{a}{0}{\sig} m_\parallel(a)\hat c(a,\om)
  \label{eq:m}
\end{equation}
adding the ideal contribution [Eq.~\eqref{eq:m0}] and the non-ideal contributions due to the active regions with $a<\sig$, where $N$ is a normalization. We cut off the integral at the particle size $\sig$ to suppress the contribution of rare large excited regions (the exact cut-off length is not relevant and modifies the absolute values of the coefficients $a_k$ below). Inserting the Taylor expansion
\begin{equation}
  \frac{m_\parallel(a)}{\mu_0} = c_1\frac{\mu_0E}{\kT} + c_3n\left(\frac{\mu_0E}{\kT}\right)^3 + \cdots
\end{equation}
with Taylor coefficients $c_k$ into Eq.~\eqref{eq:m} leads to
\begin{equation}
  \frac{m_\parallel(\om)}{\mu_0} = \sum_{k=1,3,\dots}c_k [1+G_k(\om)/N]\left(\frac{\mu_0E}{\kT}\right)^k.
  \label{eq:m:taylor}
\end{equation}
Using that $n(a)\simeq1/c(a)$, we calculate the integrals
\begin{multline}
  G_k(\om) = \IInt{a}{0}{\sig} [n(a)]^{(k-1)/2}\hat c(a,\om) \\ = \frac{\sig[c(\sig)]^{(3-k)/2}}{\al} \IInt{x}{1}{\infty} x^{(5-k)/2} e^{-\tfrac{\kap}{\om\tau_\al}x^{\al/\df}} \\ = \frac{\sig[c(\sig)]^{(3-k)/2}}{\al^2/\df} \left(\frac{\om\tau_\al}{\kap}\right)^{s_k} \Gamma(s_k,\tfrac{\kap}{\om\tau_\al})
  \label{eq:G:k}
\end{multline}
with $s_k\equiv(7-k)\df/(2\al)$ after substituting $x=(a/\sig)^{-\al}$. Here, $\Gamma(s,x)$ is the incomplete Gamma function. To derive this integral, we have used that $c(a)=c(\sig)(a/\sig)^{-\al}$ with exponent $\al\equiv\gam J(\sig)/\kTi>0$ for $T<\To$~\cite{speck19}. Moreover, we can relate the time [Eq.~\eqref{eq:tau:a}]
\begin{equation}
  \tau_1(a) = \tau_\al\left[\frac{c(\sig)}{c(a)}\right]^{\al/\df} \simeq \tau_\al(a/\sig)^{\al^2/\df}
\end{equation}
to the structural relaxation time $\tau_\al$, which leads to the second line of Eq.~\eqref{eq:G:k}.

The result $G_k(\om\tau_\al)$ only depends on $\om\tau_\al$ and temperature. It is normalized by the total number of contributing active regions
\begin{equation}
  N(\om) = \IInt{a}{0}{\infty} \hat c(a,\om) = \frac{\sig c(\sig)}{\al^2/\df}\left(\frac{\om\tau_\al}{\kap}\right)^{s_1}\Gamma(s_1)
\end{equation}
with Gamma function $\Gamma(s)$. Inserting $N(\om)$ and $G_k(\om)$ into the Taylor series Eq.~\eqref{eq:m:taylor}, we finally obtain as our central result an expression for the modulus of the $k$-th order susceptibility
\begin{equation}
  \chi_k(\om) = \frac{c_k\rho\mu_0^{k+1}}{\eps_0(\kT)^k} \frac{1+a_k(\tfrac{\om\tau_\al}{\kap})^{s_k-s_1}\Gamma(s_k,\tfrac{\kap}{\om\tau_\al})}{[1+(\om\tau_\al)^2]^{\beta/2}}.
  \label{eq:chi:k}
\end{equation}
Here we have introduced the effective strength $a_k$ of the non-ideal contributions to the susceptibilities, which scale as $a_k\propto[c(\sig)]^{(1-k)/2}$. The frequency-behavior of the non-ideal excess is non-monotonic dropping for small frequencies $\om\ll\tau_\al^{-1}$ since the correlations are erased by the larger excitations before the liquid responds to the change of the electric field.


\section{Discussion}
\label{sec:exp}

\subsection{Experimental data on glycerol}

\begin{figure}[b!]
  \centering
  \includegraphics{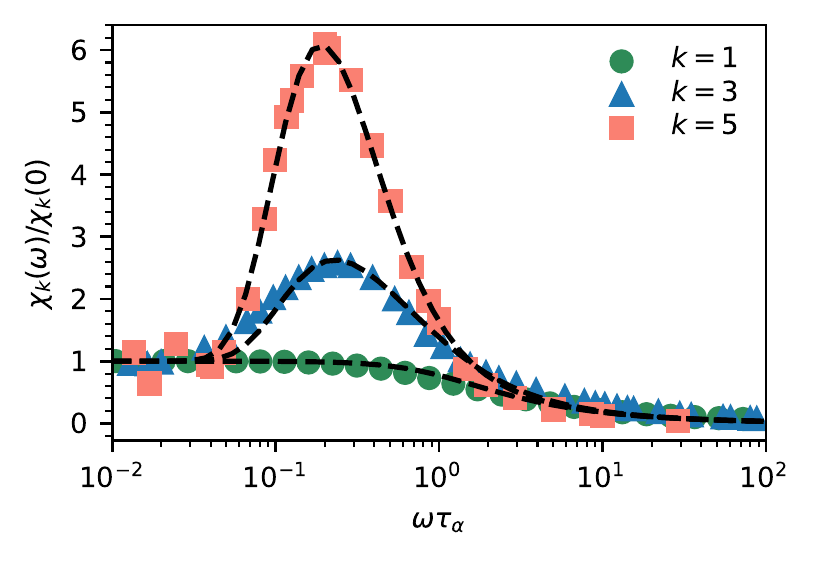}
  \caption{Reduced susceptibilities $\chi_k(\om)/\chi_k(0)$ for glycerol plotted as a function of $\om\tau_\al$ at $T=204$~K. Symbols are data from Ref.~\citenum{albe16} (Figure 1C). Dashed lines are the result of a joint fit of all three susceptibilities to Eq.~\eqref{eq:chi:k}.}
  \label{fig:gly:non}
\end{figure}

High-precision experimental results for the susceptibilities up to fifth order are presented in Ref.~\citenum{albe16} for glycerol, a molecular liquid with a dynamic glass transition temperature $\Tg\simeq185$~K~\cite{lunkenheimer02}. In Fig.~\ref{fig:gly:non}, we replot the experimental data $\chi_k(\om)/\chi_k(0)$ at $T=204$~K slightly above $\Tg$. We perform a joint fit of all three curves to Eq.~\eqref{eq:chi:k} with fit parameters $a_3$, $a_5$, $\nu\equiv\df/(2\al)$, $\kap$, and $\beta$. For the linear susceptibility we set $a_1=0$. For the exponent we obtain $\beta\simeq0.75$, which is larger than typical values obtained from simultaneously fitting the real and imaginary part of the complex linear susceptibility~\cite{lunkenheimer02}. For the coefficient of the decay time we find $\kap\simeq0.35$.

\begin{figure}[t]
  \centering
  \includegraphics{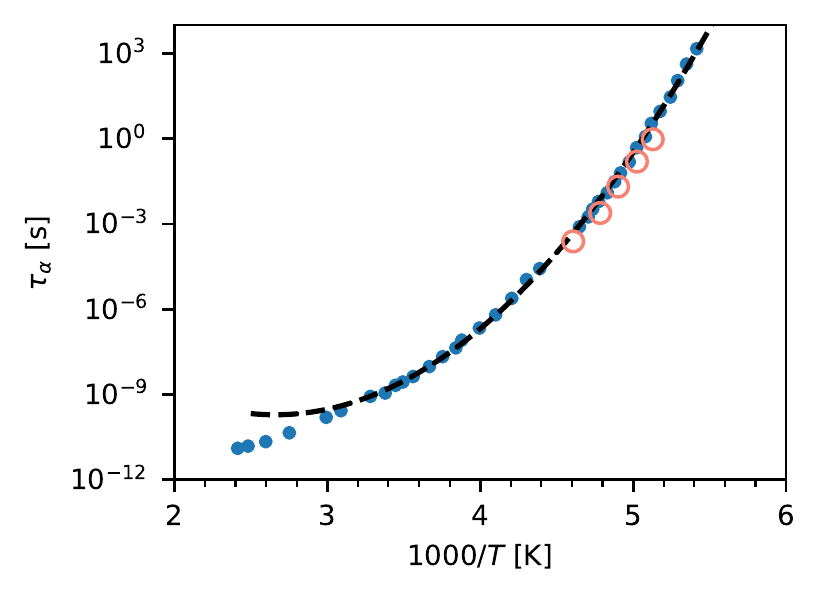}
  \caption{Structural relaxation time $\tau_\al$ as a function of inverse temperature for glycerol (closed symbols, data taken from Ref.~\citenum{lunkenheimer02}). The open circles are the relaxation times for the temperatures studied in Ref.~\citenum{albe16} (Figure 1). Dashed line is the fit to Eq.~\eqref{eq:tau:al}.}
  \label{fig:gly}
\end{figure}

Of particular interest is the value
\begin{equation}
  \nu = \frac{\df}{2\al} \simeq 0.46,
\end{equation}
which we can relate to predictions from dynamic facilitation. Employing
\begin{equation}
  \frac{\al}{\gam} = \frac{J(\sig)}{\kT_\text{o}}\left(\frac{\To}{T}-1\right)
\end{equation}
together with $J(\sig)/\kT_\text{o}=\mathcal J\sqrt{\df/\gam}$ leads to
\begin{equation}
  \nu(T) = \frac{\sqrt{\df/\gam}}{2\mathcal J}\left(\frac{\To}{T}-1\right)^{-1}.
  \label{eq:nu}
\end{equation}
To proceed, we require values for $\mathcal J$ and $\To$. Since the temperatures studied in Ref.~\citenum{albe16} span a rather narrow range, we take $\tau_\al(T)$ from Ref.~\citenum{lunkenheimer02}, which is plotted in Fig.~\ref{fig:gly} as a function of inverse temperature together with the fit to Eq.~\eqref{eq:tau:al}. The fit yields $\mathcal J\simeq 5.24$ and $\To\simeq376$~K with $\tau_\ast\simeq1.9\times10^{-10}$~s. We also plot the relaxation times inferred from Ref.~\citenum{albe16}, which are consistent but slightly smaller. Nevertheless, we will use the fitted values for $\mathcal J$ and $\To$ in the following. For $T=204$~K, we then find $\df/\gam\simeq16.76$ rearranging Eq.~\eqref{eq:nu}. Assuming a fractal dimension $\df=0.9d=2.7$ implies $\gam\simeq0.16$. In Ref.~\citenum{keys11} a number of model glass formers have been studied in computer simulations, which exhibit $0.4<\gam<0.6$ across different models. The value for $\gam$ found here is smaller, which might be attributed to the fact that glycerol is a molecular liquid in contrast to simple models based on pair potentials.

\begin{figure}[t]
  \centering
  \includegraphics{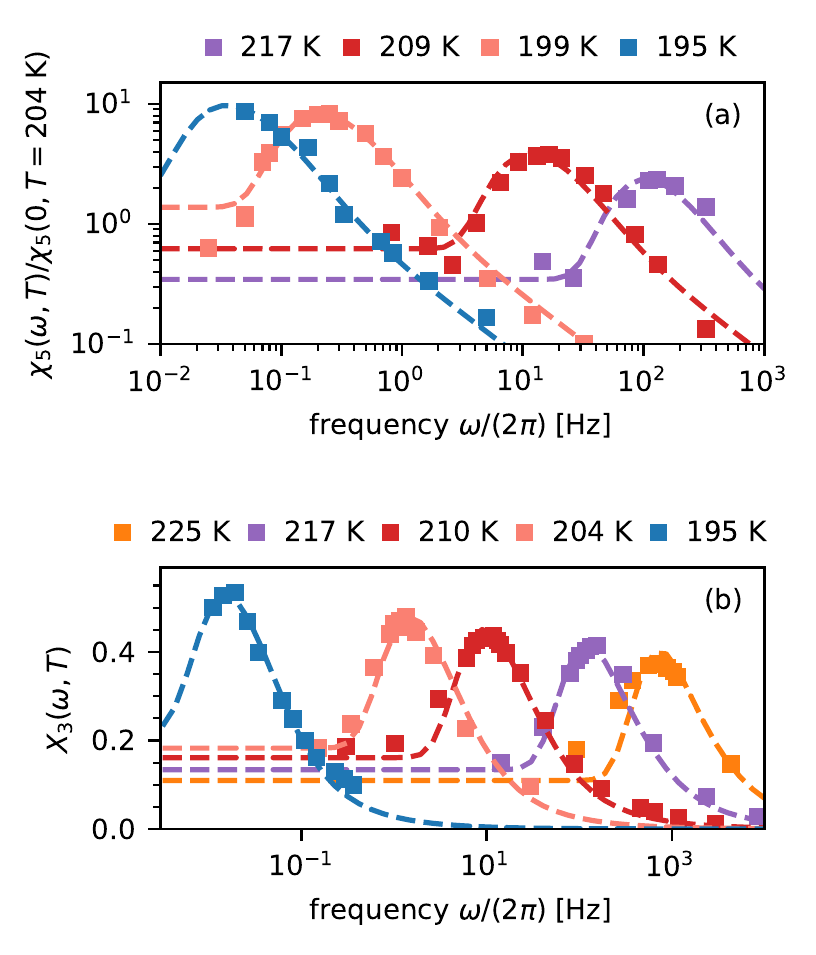}
  \caption{(a)~Fifth-order susceptibility $\chi_5(\om,T)$ for four temperatures from Ref.~\citenum{albe16}. Dashed lines show Eq.~\eqref{eq:chi:k} with the static value $\chi_5(0,T)$ as the only fit parameter. All other parameters are taken from the joint fit at $T=204$~K (Fig.~\ref{fig:gly:non}) and extrapolated using the predictions of dynamic facilitation. (b)~Reduced third-order susceptibility $X_3(\om,T)\propto\chi_3(\om,T)\times\kT[\chi_1(0,T)]^{-2}$. The data (symbols) is taken from Ref.~\citenum{crauste10}. Dashed lines show Eq.~\eqref{eq:chi:k} with the static value $\chi_3(0,T)$ as the only fit parameter analogous to (a).}
  \label{fig:gly:T}
\end{figure}

The non-linearities have fitted coefficients $a_3\simeq2.38$ and $a_5\simeq11.29$, the ratio $a_5/a_3\simeq4.74$ of which compares favorably with the prediction
\begin{equation}
  \frac{a_5}{a_3} = \frac{1}{c(\sig)} \simeq 3.98
\end{equation}
at $T=204$~K. Finally, in Fig.~\ref{fig:gly:T}(a) we show the fifth-order susceptibility $\chi_5(\om,T)$ for four temperatures other than 204~K. We determine $\nu(T)$ according to Eq.~\eqref{eq:nu} and $a_5(T)=a_5[c(\sig,204\,\text{K})/c(\sig,T)]^2$ so that the only fit parameter here is the static value $\chi_5(0,T)$. Again, we observe a very good agreement with the experimental data. The same holds for $\chi_3(\om,T)$ shown in Fig.~\ref{fig:gly:T}(b), for which we take the data provided by Ref.~\citenum{crauste10}.

\subsection{Relation between susceptibilities}

We first note that for the linear susceptibility the coefficient $a_1$ is constant and independent of temperature. Hence, $\chi_1(0)\sim1/(\kT)$ in agreement with the experimental observation. The observed simple decay of the linear susceptibility even in supercooled liquids implies $a_1\simeq0$, cf. Fig.~\ref{fig:gly:non}.

For the non-linear susceptibilities, the peak value of the non-ideal excess $\Delta\chi_k=\chi_k-\chi_k^{(0)}$ scales as
\begin{equation}
  \Delta\chi_k^\ast \sim \frac{\mu_0^{k+1}}{(\kT)^k}[c(\sig)]^{(1-k)/2}.
\end{equation}
This expression amends and refines the simplistic argument of Ref.~\citenum{speck19}. We thus find the relationship
\begin{equation}
  (\Delta\chi_3^\ast)^2 \sim \frac{\mu_0^8}{(\kT)^6}[c(\sig)]^{-2} \sim \chi_1(0)\Delta\chi_5^\ast
  \label{eq:peak}
\end{equation}
between third-order and fifth-order susceptibility. This relation has been claimed as a hallmark of RFOT in Ref.~\citenum{albe16} but is also obeyed by the expressions obtained here from arguments based on a purely dynamic relaxation mechanism.


\section{Conclusions}

We have discussed the modeling of dielectric relaxation in supercooled liquids from the perspective of dynamic facilitation~\cite{chan10}, extending the arguments that underlie the super-Arrhenius increase of the structural relaxation time to the non-linear response of a polar liquid to an external electric field. To this end, the dynamics of individual dipoles is assumed to be governed by a local amplitude $m_\parallel(a)$ that is determined using arguments from equilibrium statistical mechanics. Active localized regions sustaining mobility contribute to an enhanced response of the polarization due to dynamic correlations, the relaxation of which is governed by kinetic constraints. The predicted susceptibilities [Eq.~\eqref{eq:chi:k}] show a non-monotonic behavior as a function of frequency with a ``hump'' in very good agreement with the experimental data for glycerol. The peak height is found to grow with the concentration $c(\sig)$ of active regions (at reference length $\sig$) as $\Delta\chi_k^\ast\sim[c(\sig)]^{(1-k)/2}$.

The underlying question is whether a \emph{static} correlation length implying a collective ``rigid'' response of dipoles is necessary to rationalize the non-monotonic behavior of non-linear susceptibilities. Previous work~\cite{diez12,diez17,diez18,kim16,richert16} has already pointed out that the salient features of the non-linear response, a non-monotonic shape with a peak at $\simeq0.2\om\tau_\al$ and the temperature dependence of the peak height, can be reproduced in simple models without collective relaxation (see Ref.~\citenum{gadige18} for a reply). Moreover, the dramatic influence of microscopic dynamical rules (while leaving equilibrium quantities invariant) on the structural relaxation observed in computer simulations~\cite{santen00,nina17,ozawa19} implies that the contribution of correlated particles is strongly bound~\cite{wyart17}, which would also limit their contribution to the excess of the non-linear response. Dynamic facilitation, on the other hand, offers a perspective in which both the super-Arrhenius increase of the structural relaxation time and the excess of the non-linear response are connected to the statistics of active regions governed by kinetic constraints.


\section*{Acknowledgments}

I thank C. Patrick Royall for stimulating discussions and critically reading the manuscript. Without implying their agreement to the perspective presented here, I am grateful to Fran\c{c}ois Ladieu, Giulio Biroli, and Jean-Philippe Bouchaud for valuable feedback on the manuscript.


\section*{Data availability}

Data sharing is not applicable to this article as no new data were created in this study. Analyzed data can be downloaded from \url{https://science.sciencemag.org/content/suppl/2016/06/08/352.6291.1308.DC1} as part of the supplementary materials of Ref.~\citenum{albe16}.


%

\end{document}